\documentclass[12pt,preprint]{article}
\usepackage{amssymb}
\usepackage{amsmath}
\usepackage{graphics}
\usepackage{epsfig}

\renewcommand{\vec}[1]{{\bf #1}}
\def\bra{\langle}
\def\ket{\rangle}
\newcommand{\eqb}{\begin{equation}}
\newcommand{\eqe}{\end{equation}}
\newcommand{\dmb}{\begin{displaymath}}
\newcommand{\dme}{\end{displaymath}}
\newcommand{\pd}{\partial}

\newcommand{\eab}{\begin{eqnarray}}
\newcommand{\eae}{\end{eqnarray}}
\newcommand{\ra}{\right\rangle}
\newcommand{\la}{\left\langle}
\newcommand{\e}{\mbox{e}}
\newcommand{\be}{\begin{equation}}
\newcommand{\ee}{\end{equation}}

\begin{document}

\begin{titlepage}
\begin{flushright}
KA-TP-34-2007
\end{flushright}
\vspace{0.6cm}

\begin{center}
\Large{Correlation of energy
density\\ in deconfining SU(2) Yang-Mills thermodynamics}
\vspace{1.5cm}

\large{Jochen Keller$^\dagger$, Ralf Hofmann$^*$ and Francesco Giacosa$^{**}$}

\end{center}
\vspace{1.5cm}
\begin{center}
{\em $\mbox{}^\dagger$ Institut f\"ur Theoretische Physik\\
Universit\"at Heidelberg\\
Philosophenweg 16\\
69120 Heidelberg, Germany}
\end{center}
\vspace{1.0cm}

\begin{center}
{\em $\mbox{}^*$ Institut f\"ur Theoretische Physik\\
Universit\"at Karlsruhe (TH)\\
Kaiserstr. 12\\
76131 Karlsruhe, Germany}
\end{center}
\vspace{1.0cm}

\begin{center}
{\em $\mbox{}^{**}$ Institut f\"ur Theoretische Physik\\
Universit\"at Frankfurt\\
Max von Laue-Str. 1\\
60438 Frankfurt, Germany}
\end{center}

\vspace{1.5cm}

\begin{abstract}

We compute the two-point correlation of the energy density
for the massless mode in deconfining SU(2) Yang-Mills thermodynamics and
point towards a possible application for the physics of cold, dilute,
and stable clouds of atomic hydrogen within the Milky Way.

\end{abstract}

\end{titlepage}

\section{Introduction}

Even at high temperatures the thermodynamics of an SU(2) Yang-Mills theory is interesting judged
from both a theoretical and a phenomenological view point.

On the
theoretical side, it was argued
a long time ago that this system exhibits a magnetic infrared
instability when approached in terms of a perturbative expansion
\cite{Linde1980I,Linde1980II,
Linde1980III}. The typical alternating behavior of loop-expanded
thermodynamical quantities such as the pressure, which expresses this
fact, was impressively demonstrated in thermal
perturbation theory
\cite{Laine,Zhai,Arnold,Toimela,Kapusta,Shuryak,Chin}. In
\cite{Hofmann2005,HerbstH2004} the concept of a thermal ground state at high
temperature was introduced to address the stabilization
issue in the magnetic infrared sector. Namely, a spatial
coarse-graining over interacting, topologically nontrivial
field configurations of charge modulus unity was performed to
yield an inert adjoint scalar field and a pure-gauge configuration to
describe this ground state. The (temperature dependent) spectrum
of excitations after coarse-graining is determined by
the adjoint Higgs mechanism. In the according effective
theory this tree-level result seems to capture on the one-loop level
99.5\% of the contributions to the pressure.
The corrections are due to higher loops. It is interesting that
the presence of screened and isolated magnetic monopoles
\cite{Korthals-AltesI,Korthals-AltesII,Korthals-AltesIII}
in the system is accounted
for as a particular radiative correction to the pressure
on the two-loop level. That is, on the microsopic level
the rare dissociation of calorons or anticalorons,
largely deformed away from trivial holonomy, into pairs of massive
and screened magnetic monopoles and antimonopoles is seen as a
$\propto T^4$-correction to the pressure with a small (negative) coefficient after
spatial coarse-graining \cite{SHG2006-1}. By cutting the massless line of this two-loop
diagram the diagram determining the polarization tensor for this mode is
obtained. On shell and depending on the modulus of the spatial momentum
the polarization tensor either exhibits screening or antiscreening again thanks
to the presence of isolated, screened magnetic monopoles.

The emergence of the thermal ground state
implies the dynamical breaking of the SU(2) gauge symmetry
down to its U(1) subgroup after spatial
coarse-graining \cite{Hofmann2005,Hofmann2007}.
As a consequence, part of the spectrum of the
propagating gauge fields acquires a temperature dependent mass
thus curing the old problem of a
perturbative instability\footnote{To avoid the occurrence of infrared
  divergences in thermal perturbation theory at a fixed loop order
polarizations of lower loop order need to be resummed. The thus implied dependence of the result at a
fixed loop order on the fundamental coupling constant $g$ upsets the
naive perturbative power counting in $g$, and
the loop expansion does not converge numerically \cite{Linde1980I,Linde1980II,Linde1980III}. In contrast, the
occurrence of quasiparticle masses in the effective theory does not
allow for infrared divergences to take place: The only
potentially critical case of a three-loop diagram
containing two massless and two massive lines (two four-vertices) is
excluded by constraints on the momentum transfer in the vertices
\cite{KH2007}.} residing in the magnetic
sector \cite{Linde1980I,Linde1980II,Linde1980III}. Moreover, due to the existence
of a dynamically emerging, temperature dependent scale of
maximal resolution $|\phi|$ the computation of radiative corrections
to the free-quasiparticle situation is under control \cite{Hofmann2007,HofmannLE2006}.

On the phenomenological side,
radiative corrections in the effective theory can be used to make contact with
thermalized photon propagation. To do this the postulate is made
that an SU(2) Yang-Mills theory of scale $\sim 10^{-4}\,$eV, referred to
as SU(2)$_{\tiny\mbox{CMB}}$ in the following, underlies the U(1) gauge symmetry
of electromagnetism \cite{Hofmann2005,SHG2006-1,HofmannB2005}. If verified
experimentally through predictions based on radiative effects
\cite{SHG2006-2,SH2007,SHGS2007} then cosmological consequences would
arise \cite{GH2005}.
That is, the nontrivial ground state underlying
photon propagation would then be linked to the dynamics of an
ultralight and spatially homogeneous
scalar field (Planck-scale axion)
\cite{BardeenBellJackiwI,BardeenBellJackiwII,BardeenBellJackiwIII}
associated
with dark energy. Notice the apparent and amusing
paradox: The nonabelian gauge theory underlying the propagation of
{\sl light} would, by means of its ground state, be a
crucial ingredient in generating {\sl dark} energy\footnote{Possibly,
it is also responsible for the pressureless component so far identified
as dark
{\sl matter}, see \cite{GH2005}.}. Also, there would be immediate implications for the physics of
temperature-temperature
correlations in the cosmic microwave
background (CMB) at large angles \cite{SH2007} and, by virtue of the
connection to axion physics, a CP violating electric-magnetic cross
correlation (extracted from the according polarization map of the CMB) should be detectable by future satellite
missions. There is an alternative physical
system for which the radiative effects of SU(2)$_{\tiny\mbox{CMB}}$
would be relevant: dilute, old, and cold clouds of atomic
hydrogen in between the spiral arms of the
Milky Way \cite{BruntKnee2001}.

Notice that these results rely on
infinite-volume thermodynamics, that is, on the absence of
time-dependent sources and on the requirement that spatial boundaries, imposing
additional conditions on the gauge-field dynamics, are well farther
apart then the physical correlation lengths of the infinite-volume
situation. Away from the phase boundary at $T_c\sim 2.7\,$K, where the
mass of screened monopoles vanishes and thus the correlation length
diverges, the length scale, which separates the infinite-volume case
from the boundary case, is set by $|\phi|^{-1}=\frac{(13.87)^{3/2}}{2\pi
  T_c}\,\sqrt{T/T_c}$ \cite{Hofmann2005}. Close to $T_c$ this is
smaller but of the order of one centimeter. For this reason and for the
fact that in low-temperature condensed-matter experiments it is not the
temperature of a photon gas that is measured it is then clear
that the results of the infinite-volume approach do not apply to the
conditions prevailing in these systems. In cosmological or astrophysical situations
or in a designated black-body experiment with sufficiently large and
homogeneously thermalized volume conditions are in line with those
assumed when making a prediction for the low-frequency,
spectral distortion caused by the scattering of photons
off of isolated monopoles \cite{SHG2006-1,SHG2006-2}. In addition to a
black-body experiment for $T\ge T_c$ one may also learn about the
ground state of SU(2)$_{\tiny\mbox{CMB}}$ at lower temperatures by
detecting the onset of superconductivity in the preconfining phase associated with the
condensation of monopoles\footnote{Isolated, emerging
charges of SU(2)$_{\tiny\mbox{CMB}}$ have a dual
interpretation in the SM: What is an magnetic charge w.r.t. the
gauge fields defining the Lagrangian of SU(2)$_{\tiny\mbox{CMB}}$ is an
electric charge in the SM. Thus condensed magnetic monopoles in
SU(2)$_{\tiny\mbox{CMB}}$ are condensed electric charges in the
SM giving rise to superconductivity.}: Shortly below $T_c$
the pressure is larger than the conventional photon-gas pressure due to
the emergence of an extra degree of freedom -- a longitudinal
polarization of the photon, see also \cite{SHGS2007}. Along the same
lines a future precision survey of intergalactic
magnetic fields may be able to make quantitative contact with
SU(2)$_{\tiny\mbox{CMB}}$, see \cite{Hofmann2007} for a droplet
model describing the statistics of monopole condensing
regions.

One may also
wonder about whether standard-model (SM) weak-interaction
quantum numbers should be assigned to the
SU(2)$_{\tiny\mbox{CMB}}$ gauge fields. The point here
is that in light of the underlying gauge symmetry
the SM is a (highly successful) effective quantum
field theory with quantum numbers and interactions (Higgs sector!)
assigned to its effective fields so as to keep the theory consistent
(gauge-anomaly cancellation, renormalizability, unitarity).
On the level of fundamental gauge fields no such an
assignment is needed: The massless mode of SU(2)$_{\tiny\mbox{CMB}}$
still is the {\sl propagating} photon $\gamma$ of the SM, and local interactions with
electrically charged matter are still described through this gauge
field's mixing with the neutral gauge field of SU(2)$_W$. In addition,
no interactions
of the two massive excitations of SU(2)$_{\tiny\mbox{CMB}}$ with charged or neutral
matter and the two massive excitations of SU(2)$_W$, respectively,
takes place due to the existence of a large hierarchy in the
participating (Yang-Mills) scales on one hand and the existence of a small
compositeness scale $|\phi|$ in SU(2)$_{\tiny\mbox{CMB}}$ on the other
hand, for a more detailed discussion see \cite{GH2005}.

The purpose of the present work is to compute the two-point correlation
of the canonical energy density $\Theta_{00}$ of photons in a
thermalized gas. In the framework of SU(2)$_{\tiny\mbox{CMB}}$ the
canonical energy density of the photon gas
(massless modes in unitary gauge, see
\cite{Hofmann2005,SHG2006-1,HHR2004}) is made manifestly
SU(2) gauge invariant in the effective theory for deconfining
thermodynamics by substituting the 't Hooft tensor \cite{tTensor}
into the expression for the U(1) Belinfante energy-momentum tensor $\Theta_{\mu\nu}$. Recall, that in
unitary gauge the 't Hooft tensor reduces to the abelian field
strength for the (massless) mode pointing into the direction of the
Higgs field in the algebra. Apart from phenomenological considerations,
to compute $\la\Theta_{00}(\vec{x})\Theta_{00}(\vec{y})\ra$ is
technically interesting by itself. Namely, this quantity appears to be
accurately
calculable in terms of the one-loop photon polarization
tensor only. That is, irreducible diagrams beyond two loop appear to
yield a safely negligible contribution to $\gamma$'s polarization
tensor,
compare with the results in \cite{KH2007}. In this work we present
results in a minimal way technically to warrant a reasonably efficient flow of
the arguments. For calculational details we refer the reader to \cite{KellerDA}.

The paper is organized as follows. In Sec.\,\ref{tpc} we define the
problem in a real-time formulation of a thermalized (free) U(1) gauge
theory. Thermal and quantum parts of the correlator
$\bra\Theta_{00}(\vec{x})\Theta_{00}(\vec{y})\ket$ are calculated
analytically and compared with one another. The same program is
performed for the case of deconfining SU(2) Yang-Mills theory in
Sec.\,\ref{decsu2}. Only an estimate is possible analytically for
the vacuum contribution, and the thermal part is evaluated
numerically. We compare our results to those obtained
for the U(1) case and observe a sizable
suppression of the SU(2) correlation at low temperatures and large
distances. In Sec.\,\ref{MW} we offer a potential explanation of
why cold and dilute clouds of atomic hydrogen within the Milky Way
are apparently so stable. There is a short
summary in Sec.\,\ref{C}.

\section{Two-point correlation of energy density in thermal U(1) gauge
  theory\label{tpc}}

In this section we compute the two-point correlation of the canonical
energy density in a pure, thermalized U(1) gauge theory. Our results
will serve as a benchmark for the more involved calculation when
embedding this object into a deconfining SU(2) Yang-Mills
theory.

\subsection{General strategy}

The two-point correlation of the energy density is computed
by letting derivative operators, associated with the structure of the
energy-momentum tensor $\Theta_{\mu\nu}$, act on the real-time
propagator of the U(1) gauge field.

Recall that the traceless and symmetric (Belinfante) energy-momentum tensor of a pure
U(1) gauge theory is given as
\begin{equation}
\label{canen}
\Theta_{\mu\nu}=-F_{\mu}\!^{\lambda}F_{\nu\lambda}+\frac{1}{4}g_{\mu\nu}F^{\kappa\lambda}F_{\kappa\lambda}\,,
\end{equation}
where $F_{\mu\nu}\equiv\pd_\mu A_\nu-\pd_\nu A_\mu$, and $A_\mu$ denotes the
U(1) gauge field. In the deconfining phase of SU(2) Yang-Mills
thermodynamics the effective theory contains an inert, adjoint Higgs
field $\phi$ \cite{Hofmann2005,Hofmann2007}. The field $\phi$ emerges
by a spatial coarse-graining process over topologically nontrivial field
configurations of trivial holonomy: Harrington-Shepard calorons and
anticalorons \cite{HS}. These nonpropagating, BPS saturated, periodic solutions to the
euclidean Yang-Mills equations are deformed by gluon
exchanges leading to short-lived nontrivial holonomy and
thus short-lived, separated magnetic charges
\cite{NahmP,NahmL,LeeLu1998,KraanVanBaalNPB1998-1,KraanVanBaalNPB1998-2,KraanVanBaalNPB1998-3,Brower1998}.
This situation is described by a pure-gauge configuration in the
effective theory.

After spatial coarse-graining and owing to the presence of the field $\phi$,
the field strength
$F_{\mu\nu}$ of the abelian theory in Eq.\,(\ref{canen}) can be replaced by the 't Hooft
tensor ${\cal F}_{\mu\nu}$ to define an SU(2) gauge invariant
energy-momentum tensor. One has \cite{tTensor}
\begin{equation}
\label{'tHooft}
{\cal F}_{\mu\nu}\equiv \frac{1}{|\phi|}\phi_a
G_{\mu\nu}^a-\frac{1}{e |\phi|^3}\epsilon^{abc}\phi_a(D_\mu\phi)_b(D_\nu\phi)_c\,,
\end{equation}
where $G^a_{\mu\nu}$ is the SU(2) field strength of topologically
trivial, coarse-grained fluctuations, $D_\mu$ denotes the adjoint
covariant derivative, and $e$ is the effective gauge
coupling. Obviously, the quantity defined by the right-hand side of
Eq.\,(\ref{'tHooft}) is SU(2) gauge invariant. In
unitary gauge $\phi_a=\delta_{a3}|\phi|$ the 't Hooft tensor reduces to the
abelian tensor $F_{\mu\nu}$ defined on the massless gauge field
$A_\mu^3$.

We are interested in computing the connected correlation
function $\la\Theta_{00}(x)\Theta_{00}(y)\ra$ in four-dimensional
Minkowskian spacetime ($g_{00}=1$). This is done
by applying Wick's theorem to express
$\la\Theta_{00}(x)\Theta_{00}(y)\ra$ in terms
of the propagator $D_{\mu\nu}$, which in Coulomb gauge and momentum space is given
as
\begin{eqnarray}
\label{U1prop}
D_{\mu\nu}(p,T)&=&D^{\tiny\mbox{vac}}_{\mu\nu}(p,T)+D^{\tiny\mbox{th}}_{\mu\nu}(p,T)\nonumber\\
&=&-P^T_{\mu\nu}(p)\frac{i}{p^2+i\epsilon}+i\frac{u_\mu
  u_\nu}{\vec{p}^2}-P^T_{\mu\nu}(p)2\pi\delta(p^2)n_B(\beta|p_0|)\,,
\end{eqnarray}
where $u_\mu=(1,0,0,0)$, $\beta\equiv \frac{1}{T}$,
\begin{eqnarray}
P^T_{00}(p)&\equiv&P^T_{0i}(p)=P^T_{i0}(p)=0,\nonumber\\
P^T_{ij}(p)&\equiv&\delta_{ij}-\frac{p_i p_j}{\vec{p}^2}\,,
\end{eqnarray}
and $n_B(x)\equiv\frac{1}{e^x-1}$. By virtue of Eq.\,(\ref{canen}) one
obtains
\begin{eqnarray}
\label{formalexpTheta00}
\bra\Theta_{00}(x)\Theta_{00}(y)\ket &=&
    2\bra\partial_{x^0}A^{\lambda}(x)\partial_{y^0}A^{\tau}(y)\ket
    \bra\partial_{x^0}A_{\lambda}(x)\partial_{y^0}A_{\tau}(y)\ket\nonumber\\
&&-g_{00}\bra\partial_{x^0}A^{\lambda}(x)\partial_{y^{\sigma}}A^{\tau}(y)\ket
    \bra\partial_{x^0}A_{\lambda}(x)\partial_{y_\sigma}A_{\tau}(y)\ket\nonumber\\
&&+g_{00}\bra\partial_{x^0}A^{\lambda}(x)\partial_{y_\sigma}A^{\tau}(y)\ket
    \bra\partial_{x^0}A_{\lambda}(x)\partial_{y^\tau}A_{\sigma}(y)\ket\nonumber\\
&&-g_{00}\bra\partial_{x_\kappa}A^{\lambda}(x)\partial_{y^0}A^{\tau}(y)\ket
    \bra\partial_{x^\kappa}A_{\lambda}(x)\partial_{y^0}A_{\tau}(y)\ket\nonumber\\
&&+g_{00}\bra\partial_{x_\kappa}A^{\lambda}(x)\partial_{y^0}A^{\tau}(y)\ket
    \bra\partial_{x^\lambda}A_{\kappa}(x)\partial_{y^0}A_{\tau}(y)\ket\nonumber\\
&&+\frac{g_{00}^2}{2}
    \bra\partial_{x_\kappa}A^{\lambda}(x)\partial_{y_\sigma}A^{\tau}(y)\ket
    \bra\partial_{x^\kappa}A_{\lambda}(x)\partial_{y^\sigma}A_{\tau}(y)\ket\nonumber\\
&&-\frac{g_{00}^2}{2}
    \bra\partial_{x_\kappa}A^{\lambda}(x)\partial_{y_\sigma}A^{\tau}(y)\ket
    \bra\partial_{x^\kappa}A_{\lambda}(x)\partial_{y^\tau}A_{\sigma}(y)\ket\nonumber\\
&&-\frac{g_{00}^2}{2}
    \bra\partial_{x_\kappa}A^{\lambda}(x)\partial_{y_\sigma}A^{\tau}(y)\ket
    \bra\partial_{x^\lambda}A_{\kappa}(x)\partial_{y^\sigma}A_{\tau}(y)\ket\nonumber\\
&&+\frac{g_{00}^2}{2}
    \bra\partial_{x_\kappa}A^{\lambda}(x)\partial_{y_\tau}A^{\sigma}(y)\ket
    \bra\partial_{x^\lambda}A_{\kappa}(x)\partial_{y^\sigma}A_{\tau}(y)\ket\nonumber\\
&&+2\bra\partial_{x^0}A^0(x)\partial_{y^0}A^0(y)\ket
    \bra\partial_{x^0}A_0(x)\partial_{y^0}A_0(y)\ket\nonumber\\
&&-2\bra\partial_{x^0}A^0(x)\partial_{y^\tau}A^0(y)\ket
    \bra\partial_{x^0}A_0(x)\partial_{y_\tau}A_0(y)\ket\nonumber\\
&&+2\bra\partial_{x^\tau}A^0(x)\partial_{y^\sigma}A^0(y)\ket
    \bra\partial_{x_\tau}A_0(x)\partial_{y_\sigma}A_0(y)\ket\nonumber\\
&&-2g_{00}\bra\partial_{x^0}A^0(x)\partial_{y^0}A^0(y)\ket
    \bra\partial_{x^0}A_0(x)\partial_{y^0}A_0(y)\ket\nonumber\\
&&+4g_{00}\bra\partial_{x^0}A^0(x)\partial_{y^\tau}A^0(y)\ket
    \bra\partial_{x^0}A_0(x)\partial_{y_\tau}A_0(y)\ket\nonumber\\
&&-2g_{00}\bra\partial_{x^\tau}A^0(x)\partial_{y^\sigma}A^0(y)\ket
    \bra\partial_{x_\tau}A_0(x)\partial_{y_\sigma}A_0(y)\ket\nonumber\\
&&+\frac{g_{00}^2}{2}\bra\partial_{x^0}A^0(x)\partial_{y^0}A^0(y)\ket
    \bra\partial_{x^0}A_0(x)\partial_{y^0}A_0(y)\ket\nonumber\\
&&-g_{00}^2\bra\partial_{x^0}A^0(x)\partial_{y^\tau}A^0(y)\ket
    \bra\partial_{x^0}A_0(x)\partial_{y_\tau}A_0(y)\ket\nonumber\\
&&+\frac{g_{00}^2}{2}\bra\partial_{x^\tau}A^0(x)\partial_{y^\sigma}A^0(y)\ket
    \bra\partial_{x_\tau}A_0(x)\partial_{y_\sigma}A_0(y)\ket\,.\nonumber\\
\end{eqnarray}
Notice that at this stage ambiguities related to the various
possibilities of time ordering in the two-point function of the gauge
field $A_\mu$ cancel out. The last nine lines in Eq.\,(\ref{formalexpTheta00}) arise
from the term $\propto u_\mu u_\nu$ in the propagator, see
Eq.\,(\ref{U1prop}).

\subsection{Real-time formalism: Decomposition into thermal and vacuum parts}

\begin{figure}
\begin{center}
\vspace{5.3cm}
\includegraphics{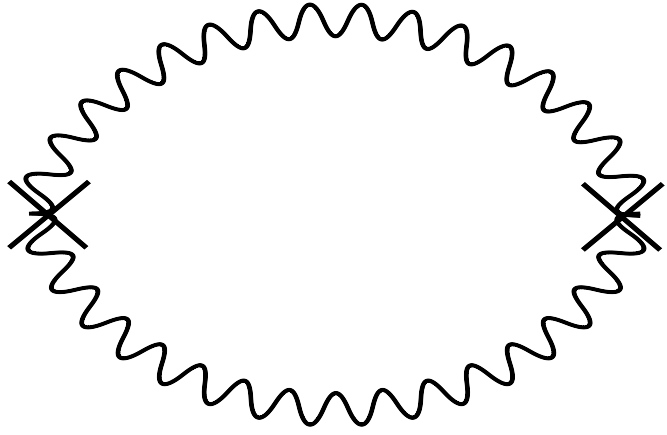}
\end{center}
\caption{Feynman diagram for the correlator
  $\bra\Theta_{00}(x)\Theta_{00}(y)\ket$ in a pure U(1) gauge
theory. Crosses denote the insertion of the local, composite operator $\Theta_{00}$.\label{Fig-1}}
\end{figure}

In evaluating the expression in Eq.\,(\ref{formalexpTheta00})
the derivative operators are taken out of the expectation, and
Eq.\,(\ref{U1prop}) is used. By momentum conservation the expression
Eq.\,(\ref{formalexpTheta00}) separates into purely thermal and
purely vacuum contributions, see Fig.\,\ref{Fig-1}.
Using
\begin{eqnarray}
\label{contrthermu1}
P^{T\,\lambda\tau}(p)P^{T}_{\lambda\tau}(k)
&=&1+\frac{(\vec{p}\cdot
  \vec{k})^2}{\vec{p}^2\vec{k}^2}\,,\nonumber\\
P^{T}_{\lambda\sigma}(p)k^\lambda
k^\sigma&=&\vec{k}^2-\frac{(\vec{p}\cdot
  \vec{k})^2}{\vec{p}^2}\,,\nonumber\\
P^{T\,\kappa\tau}(p)P^{T}_{\kappa\sigma}(k)p^\sigma k_\tau
&=&-(\vec{p}\cdot \vec{k})
\left(\frac{(\vec{p}\cdot \vec{k})^2}{\vec{p}^2 \vec{k}^2}-1\right)
\end{eqnarray}
the purely thermal contribution
$\bra\Theta_{00}(x)\Theta_{00}(y)\ket^{\tiny\mbox{th}}$
(performing the trivial integration over the 0-components of the
momenta) reads
\begin{eqnarray}
\label{thermcalcu1}
\bra\Theta_{00}(\vec{x})\Theta_{00}(\vec{y})\ket^{\tiny\mbox{th}}&=&
        \left(\int\!\!\frac{d^3p}{(2\pi)^3}
        |\vec{p}|\,n_B(\beta|\vec{p}|)
        \,\e^{i\vec{p}\vec{z}}\right)^2\nonumber\\
&+&\int\!\!\frac{d^3p}{(2\pi)^3}\int\!\!\frac{d^3k}{(2\pi)^3}
        \left(\frac{\vec{p}\vec{k}}{|\vec{p}||\vec{k}|}\right)^2
        |\vec{p}||\vec{k}|\,n_B(\beta|\vec{p}|)n_B(\beta|\vec{k}|)
        \,\e^{i\vec{p}\vec{z}}\,\e^{i\vec{k}\vec{z}}\nonumber\\
\end{eqnarray}
where $\vec{z}\equiv \vec{x}-\vec{y}$. In deriving
Eq.\,(\ref{thermcalcu1}) we have set $x^0=y^0$ thus neglecting
oscillatory terms. This prescription should reflect the time-averaged
energy transport between points $\vec{x}$ and $\vec{y}$ and is technically much easier to handle.

Inserting the vacuum part of the propagator in Eq.\,(\ref{U1prop}) into
Eq.\,(\ref{formalexpTheta00}), performing similar contractions as in
Eq.\,(\ref{contrthermu1}), and rotating to euclidean signature
($p_0,k_0\to ip_0,ik_0$, $x^0,y^0\to -ix^0,-iy^0$, $g_{\mu\nu}\to -\delta_{\mu\nu}$), we obtain
\begin{eqnarray}
\label{thetathetavacu1}
&&\bra\Theta_{00}(x)\Theta_{00}(y)\ket^{\tiny\mbox{vac}}\nonumber\\
&&=\frac{9}{2}\left(\int\frac{d^4p}{(2\pi)^4}
   \,p_{0}^2\,\frac{\e^{ip\zeta}}{p^2}\right)^2
  +\frac{1}{2}\left(\int\frac{d^4p}{(2\pi)^4}
   \,|\vec{p}|^2\,\frac{\e^{ip\zeta}}{p^2}\right)^2\nonumber\\
&&+6\int\frac{d^4p}{(2\pi)^4}\int\frac{d^4k}{(2\pi)^4}
   \left(\frac{\vec{p}\vec{k}}{|\vec{p}||\vec{k}|}\right)
   p_0 k_0 |\vec{p}| |\vec{k}|
   \,\frac{\e^{ip\zeta}}{p^2}\,\frac{\e^{ik\zeta}}{k^2}\nonumber\\
&&+\frac{1}{2}\int\frac{d^4p}{(2\pi)^4}\int\frac{d^4k}{(2\pi)^4}
   \left(\frac{\vec{p}\vec{k}}{|\vec{p}||\vec{k}|}\right)^2
   \left(9\,p_0^2 k_0^2 + |\vec{p}|^2 |\vec{k}|^2\right)
   \,\frac{\e^{ip\zeta}}{p^2}\,\frac{\e^{ik\zeta}}{k^2}\nonumber\\
&&+2\left(\int\frac{d^4p}{(2\pi)^4}\,p_{0}^2
   \,\frac{\e^{ip\zeta}}{\vec{p}^2}\right)^2\nonumber\\
&&+2\int\frac{d^4p}{(2\pi)^4}\int\frac{d^4k}{(2\pi)^4}
   \left(\frac{\vec{p}\vec{k}}{|\vec{p}||\vec{k}|}\right)
   p_0 k_0 |\vec{p}| |\vec{k}|
   \frac{\e^{ip\zeta}}{\vec{p}^2}\,\frac{\e^{ik\zeta}}{\vec{k}^2}\nonumber\\
&&+\frac{9}{2}\int\frac{d^4p}{(2\pi)^4}\int\frac{d^4k}{(2\pi)^4}
   \left(\frac{\vec{p}\vec{k}}{|\vec{p}||\vec{k}|}\right)^2
   |\vec{p}|^2 |\vec{k}|^2
   \,\frac{\e^{ip\zeta}}{\vec{p}^2}\,\frac{\e^{ik\zeta}}{\vec{k}^2}\,.\nonumber\\
\end{eqnarray}
where $\zeta\equiv x-y$ and $p\zeta=p_\mu\zeta_\mu, k\zeta=k_\mu\zeta_\mu,
p^2=p_\mu p_\mu,$ and $k^2=k_\mu k_\mu$. The last three lines in
Eq.\,(\ref{thetathetavacu1}) arise
from the term $\propto u_\mu u_\nu$ in the propagator, see Eq.\,(\ref{U1prop}).

\subsection{Results}

To evaluate the integrals in Eqs.\,(\ref{thermcalcu1}) and
(\ref{thetathetavacu1}) we introduce rescaled momenta
$\tilde{p}_\mu\equiv \beta p_\mu$ and $\tilde{k}_\mu\equiv\beta k_\mu$. The integrals in Eqs.\,(\ref{thermcalcu1}) and
(\ref{thetathetavacu1}) are now expressed in terms of 3D and 4D
spherical coordinates, respectively, and the integration over azimuthal
angles is performed in a straight-forward way in both cases.
As a result, the integrals over the remaining variables
factorize for each term in Eqs.\,\,(\ref{thermcalcu1}) and
(\ref{thetathetavacu1}).

In case of
$\bra\Theta_{00}(\textbf{x})\Theta_{00}(\textbf{y})\ket^{\tiny\mbox{th}}$
we choose $\vec{z}$ to point into the 3-direction. Making use of the
spherical expansion of the exponential
\begin{equation}
\label{spher3D}
e^{i|\vec{\tilde{q}}||\vec{\tilde{\vec{z}}}|\cos\theta}=
\sum\limits_{l=0}^{\infty} i^l (2l+1) j_l(|\vec{\tilde{q}}||\vec{\tilde{\vec{z}}}|) P_l(\cos\theta)\,,
\end{equation}
where $\theta\equiv\angle({\vec{q},\vec{\zeta}})$,
$\vec{\tilde{\vec{z}}}\equiv\frac{\vec{z}}{\beta}$, $q=p,k$, and $j_l$
denotes a spherical Bessel function, expressing
polynomial factors in $\cos\theta$ in terms of linear combinations of
Legendre polynomials $P_l(\cos\theta)$, and exploiting their
orthonormality relation
\begin{equation}
\label{orthnorm}
\int\limits_{-1}^{+1}dx \,P_n(x) P_m(x)=\frac{2}{2m+1}\delta_{mn}\,,\ \ \
(m,n\ \ \mbox{integer})\,,
\end{equation}
in integrating over the polar angle $\theta$, we arrive at
\begin{eqnarray}
\label{contrthermu1cal}
\bra\Theta_{00}(\vec{x})\Theta_{00}(\vec{y})\ket^{\tiny\mbox{th}}
&=&\frac{1}{(2\pi)^6\beta^8}\left(\frac{64\pi^2}{3}
                     \left(\int\limits_0^{\infty}d|\tilde{\vec{p}}|\frac{|\tilde{\vec{p}}|^3}
                             {\e^{|\tilde{\vec{p}}|}-1} j_0(|\tilde{\vec{z}}||\tilde{\vec{p}}|)\right)^2\right.\nonumber\\
& &\left.+\frac{32\pi^2}{3}
                     \left(\int\limits_0^{\infty}d|\tilde{\vec{p}}|\frac{|\tilde{\vec{p}}|^3}
                             {\e^{|\tilde{\vec{p}}|}-1} j_2(|\tilde{\vec{z}}||\tilde{\vec{p}}|)\right)^2\right)\nonumber\\
\end{eqnarray}
Performing the integration over $|\tilde{\vec{p}}|$, we have \cite{Gradshteyn}
\begin{eqnarray}
\label{fru1therm}
&&\bra\Theta_{00}(\vec{x})\Theta_{00}(\vec{y})\ket^{\tiny\mbox{th}}=\nonumber\\
&&\frac{1}{(2\pi)^6\beta^8}\Bigg(\frac{64\pi^2}{3}\left(\frac{1}{|\tilde{\vec{z}}|^4}
     -\frac{\pi^3\coth(\pi|\tilde{\vec{z}}|)\textrm{cosech}^2(\pi|\tilde{\vec{z}}|)}
     {|\tilde{\vec{z}}|}\right)^2\nonumber\\
&&\left.+\frac{32\pi^2}{3}\left(\frac{-8+\pi|\tilde{\vec{z}}|(3\coth(\pi|\tilde{\vec{z}}|)
     +\pi|\tilde{\vec{z}}|(3+2\pi|\tilde{\vec{z}}|\coth(\pi|\tilde{\vec{z}}|))
     \,\textrm{cosech}^2(\pi|\tilde{\vec{z}}|))}
     {2|\tilde{\vec{z}}|^4}\right)^2\right)\nonumber\\
\end{eqnarray}
\begin{figure}
\begin{center}
\vspace{3.3cm}
\includegraphics{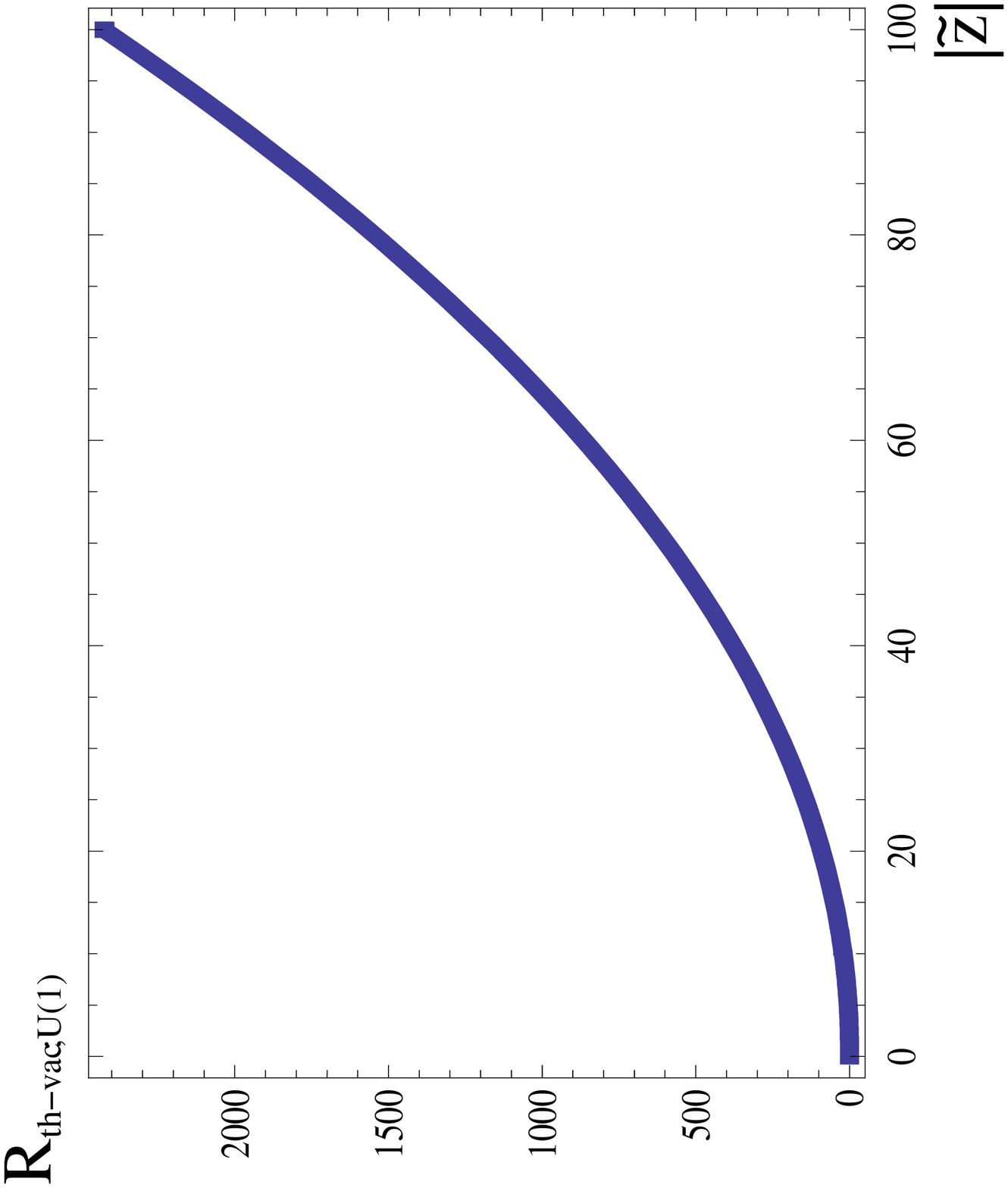}
\end{center}
\caption{The ratio
$R_{\tiny\mbox{th}-\tiny\mbox{vac};\tiny\mbox{U(1)}}\equiv\frac{\bra\Theta_{00}(\vec{x})\Theta_{00}(\vec{y})\ket^{\tiny\mbox{th}}}
{\bra\Theta_{00}(\vec{x})\Theta_{00}(\vec{y})\ket^{\tiny\mbox{vac}}}$
  as a function of $|\tilde{\vec{z}}|$ for a thermalized pure U(1) gauge theory.\label{Fig-2}}
\end{figure}
In case of
$\bra\Theta_{00}(x)\Theta_{00}(y)\ket^{\tiny\mbox{vac}}$
we restrict to $x_0=y_0$ to be able to compare with
$\bra\Theta_{00}(\textbf{x})\Theta_{00}(\textbf{y})\ket^{\tiny\mbox{th}}$.
Furthermore we chose $\vec{z}$ to point into
the 3-direction. We expect $\bra\Theta_{00}(x)\Theta_{00}(y)\ket^{\tiny\mbox{vac}}$ to be
a negligible correction to $\bra\Theta_{00}(\textbf{x})\Theta_{00}(\textbf{y})\ket^{\tiny\mbox{th}}$
for $|\tilde{\vec{z}}|>1$.
In analogy to
the thermal case each term in Eq.\,(\ref{thetathetavacu1}) factorizes
upon azimuthal integration. Making use of the
spherical expansion of the exponential
\begin{equation}
\label{spher4D}
e^{i|\vec{\tilde{q}}||\vec{\tilde{\vec{z}}}|\sin\psi\,\cos\theta}=
\sum\limits_{l=0}^{\infty} i^l (2l+1) j_l(|\vec{\tilde{q}}||\vec{\tilde{\vec{z}}}|\sin\psi) P_l(\cos\theta)\,,
\end{equation}
where $\psi$ is the second polar angle, expressing
polynomial factors in $\cos\theta$ in terms of linear combinations of
Legendre polynomials $P_l(\cos\theta)$, and exploiting their
orthonormality relation (\ref{orthnorm}) in integrating over the first
polar angle $\theta$, we arrive at
\begin{eqnarray}
\label{before2polvacu1}
&&\bra\Theta_{00}(\vec{x})\Theta_{00}(\vec{y})\ket^{\tiny\mbox{vac}}
     =\frac{1}{(2\pi)^8\beta^8}\left(96\pi^2
     \left(\int\limits_0^{\infty}\!\!d|\tilde{p}|\,|\tilde{p}|^3
     \int\limits_0^{\pi}\!\!d\psi\,\sin^2\psi \cos^2\psi
     j_0(|\tilde{p}||\tilde{\vec{z}}|\sin\psi) \right)^2\right. \nonumber\\
&&+\left.48\pi^2
     \left(\int\limits_0^{\infty}\!\!d|\tilde{p}|\,|\tilde{p}|^3
     \int\limits_0^{\pi}\!\!d\psi\,\sin^2\psi \cos^2\psi j_2(|\tilde{p}||\tilde{\vec{z}}|\sin\psi) \right)^2\right. \nonumber\\
&&+\left.96\pi^2
     \left(\int\limits_0^{\infty}\!\!d|\tilde{p}|\,|\tilde{p}|^3
     \int\limits_0^{\pi}\!\!d\psi\,\sin^3\psi \cos\psi
     j_1(|\tilde{p}||\tilde{\vec{z}}|\sin\psi) \right)^2 \nonumber\right. \\
&&+\left.\frac{32\pi^2}{3}
     \left(\int\limits_0^{\infty}\!\!d|\tilde{p}|\,|\tilde{p}|^3
     \int\limits_0^{\pi}\!\!d\psi\,\sin^4\psi j_0(|\tilde{p}||\tilde{\vec{z}}|\sin\psi) \right)^2\right.  \nonumber\\
&&+\left.\frac{16\pi^2}{3}
     \left(\int\limits_0^{\infty}\!\!d|\tilde{p}|\,|\tilde{p}|^3
     \int\limits_0^{\pi}\!\!d\psi\,\sin^4\psi j_2(|\tilde{p}||\tilde{\vec{z}}|\sin\psi) \right)^2\right.  \nonumber\\
&&+\left.32\pi^2
     \left(\int\limits_0^{\infty}\!\!d|\tilde{p}|\,|\tilde{p}|^3
     \int\limits_0^{\pi}\!\!d\psi\,\cos^2\psi j_0(|\tilde{p}||\tilde{\vec{z}}|\sin\psi) \right)^2\right.  \nonumber\\
&&+\left.32\pi^2
     \left(\int\limits_0^{\infty}\!\!d|\tilde{p}|\,|\tilde{p}|^3
     \int\limits_0^{\pi}\!\!d\psi\,\sin\psi \cos\psi j_1(|\tilde{p}||\tilde{\vec{z}}|\sin\psi) \right)^2\right.\nonumber\\
&&+\left.24\pi^2
     \left(\int\limits_0^{\infty}\!\!d|\tilde{p}|\,|\tilde{p}|^3
     \int\limits_0^{\pi}\!\!d\psi\,\sin^2\psi j_0(|\tilde{p}||\tilde{\vec{z}}|\sin\psi) \right)^2\right.\nonumber\\
&&+\left.48\pi^2
     \left(\int\limits_0^{\infty}\!\!d|\tilde{p}|\,|\tilde{p}|^3
     \int\limits_0^{\pi}\!\!d\psi\,\sin^2\psi j_2(|\tilde{p}||\tilde{\vec{z}}|\sin\psi)\right)^2\right)\,,\nonumber\\
&&
\end{eqnarray}
where now $|\tilde{p}|\equiv
\sqrt{\tilde{p}_0^2+\tilde{p}_1^2+\tilde{p}_2^2+\tilde{p}_3^2}$,
and $j_0, j_1, j_2$ are spherical Bessel functions. In Eq.\,(\ref{before2polvacu1}) the last four lines arise
from the term $\propto u_\mu u_\nu$ in the propagator, see
Eq.\,(\ref{U1prop}). They vanish because no
energy transfer between points $\vec{x}$ and $\vec{y}$ is mediated
by the Coulomb part of the photon propagator. Upon performing the
integration over $\psi$ the third and seventh line 
vanish for symmetry reasons. Our final result is
\begin{eqnarray}
\label{finu1vac}
\bra\Theta_{00}(\vec{x})\Theta_{00}(\vec{y})\ket^{\tiny\mbox{vac}}
&=&\frac{1}{(2\pi)^8\beta^8}\cdot96\pi^2\cdot
   \left(\frac{2\pi}{|\tilde{\vec{z}}|^4}\right)^2
   +\frac{1}{(2\pi)^8\beta^8}\cdot48\pi^2\cdot
   \left(\frac{-8\pi}{|\tilde{\vec{z}}|^4}\right)^2\nonumber\\
&&+\frac{1}{(2\pi)^8\beta^8}\cdot\frac{32\pi^2}{3}\cdot
   \left(\frac{-2\pi}{|\tilde{\vec{z}}|^4}\right)^2
   +\frac{1}{(2\pi)^8\beta^8}\cdot\frac{16\pi^2}{3}\cdot
   \left(\frac{8\pi}{|\tilde{\vec{z}}|^4}\right)^2\nonumber\\
&=&\frac{15}{\pi^4\beta^8\,|\tilde{{\vec{z}}}|^8}
  =\frac{0.15399}{|\vec{z}|^8}\,.
\end{eqnarray}
We have checked this result by a position-space calculation in Feynman
gauge where the vacuum part of the propagator is given as
\eqb
\label{feynmanprop}
D_{\mu\nu}^{\tiny\mbox{vac}}(x)=\frac{1}{4\pi^2 x^2}g_{\mu\nu}\,.
\eqe
In Fig.\,\ref{Fig-2} the ratio
$R_{\tiny\mbox{th}-\tiny\mbox{vac};\tiny\mbox{U(1)}}
\equiv\frac{\bra\Theta_{00}(\vec{x})\Theta_{00}(\vec{y})\ket^{\tiny\mbox{th}}}
{\bra\Theta_{00}(\vec{x})\Theta_{00}(\vec{y})\ket^{\tiny\mbox{vac}}}$
  is shown as a function of $|\tilde{\vec{z}}|$. For example, at $T=5.5, 8.2,
  10.9\,$K a distance $|\vec{z}|$ of 1\,cm corresponds to
  $|\tilde{\vec{z}}|\sim 24, 36, 48$, respectively. As a consequence, the
  thermal part of $\bra\Theta_{00}(\vec{x})\Theta_{00}(\vec{y})\ket$
  dominates the vacuum part by at least a factor of hundred.

\section{The case of deconfining thermal SU(2) gauge
  theory\label{decsu2}}

In this section we consider the massless mode surviving the dynamical gauge symmetry
breaking SU(2)$\to$U(1) in the deconfining phase of SU(2) Yang-Mills
thermodynamics. This symmetry breaking is a consequence of the
nontrivial thermal ground state composed of interacting calorons and
anticalorons. Upon a unique spatial coarse-graining this ground
state is described by spatially homogeneous field configurations
\cite{Hofmann2005}, and two out of three directions in the SU(2)
algebra dynamically acquire a temperature dependent mass. Working in unitary-Coulomb gauge, where the adjoint
Higgs field $\phi^a$ is given as  $\phi^a=\delta^{a3}|\phi|$
($a=1,2,3$), the tree-level massless, coarse-grained, topologically
trivial gauge field is $A_\mu^3$. Our goal is to obtain a measure
for the energy transfer between points $\vec{x}$ and $\vec{y}$ as
mediated by this mode when interacting with the
two massive excitations. This energy transfer is characterized by the
two-point correlator
$\bra\Theta_{00}(\textbf{x})\Theta_{00}(\textbf{y})\ket^{\tiny\mbox{th}}$
where $\Theta_{00}$ is now calculated as in
Eq.\,(\ref{canen}) replacing\footnote{This can be made manifestly SU(2)
  gauge invariant by substituting the 't Hooft tensor \cite{tTensor} for the field
  strength into Eq.\,(\ref{canen}).} $A_\mu$ by $A_\mu^3$.

\subsection{Radiative modification of dispersion law for massless mode\label{RMd}}

\begin{figure}
\begin{center}
\vspace{6.0cm}
\includegraphics{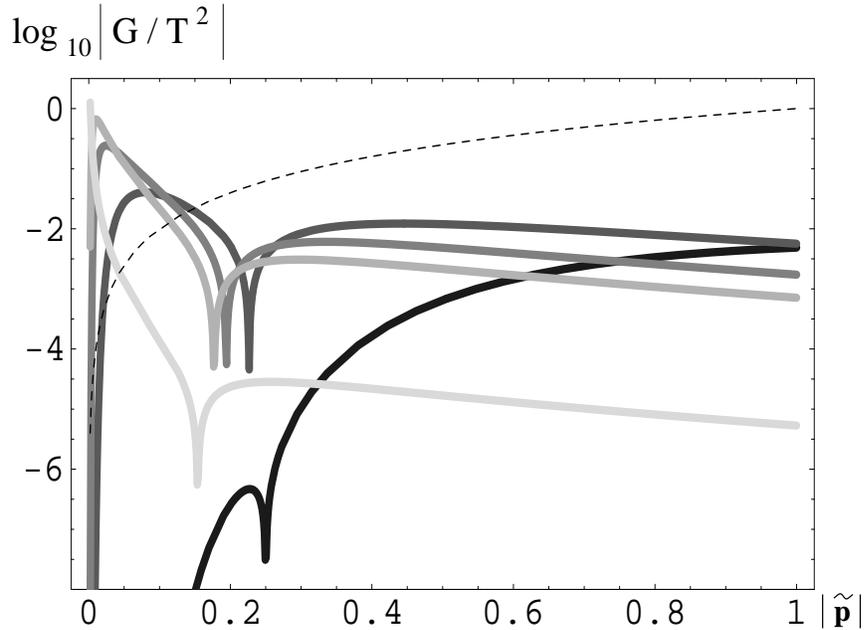}
\caption{\protect{\label{Fig-3}}$\log_{10}\left|\frac{G}{T^2}\right|$ as
  a function of
$|\tilde{\vec{p}}|$
for $\lambda=1.12\,\lambda_{c}$ (black), $\lambda=2\,\lambda_{c}$ (dark grey),
$\lambda=3\,\lambda_{c}$ (grey), $\lambda=4\,\lambda_{c}$ (light grey), $\lambda=20\,\lambda_{c}$
(very light grey). This result is obtained by appealing to the
approximation $\tilde{p}^2=0$. The full calculation shows similar results for
finite $|\tilde{\vec{p}}|$. However, there we have
$\lim_{|\tilde{\vec{p}}|\to 0}\left|\frac{G}{T^2}\right|>0$ in contrast to
the here-indicated result. The dashed curve is a
plot of the function
$f(|\tilde{\vec{p}}|)=2\log_{10}|\tilde{\vec{p}}|$. Here $\lambda\equiv
13.87\frac{T}{T_c}=\frac{2\pi T}{\Lambda}$. Photons are strongly screened at
$|\tilde{\vec{p}}|$-values for which $\log_{10}\left|\frac{G}{T^2}\right|>f(|\tilde{\vec{p}}|)$, that is, to the left
of the dashed line. The dips correspond to the zeros of $G$.}
\end{center}
\end{figure}

\begin{figure}
\begin{center}
\vspace{5.3cm}
\includegraphics{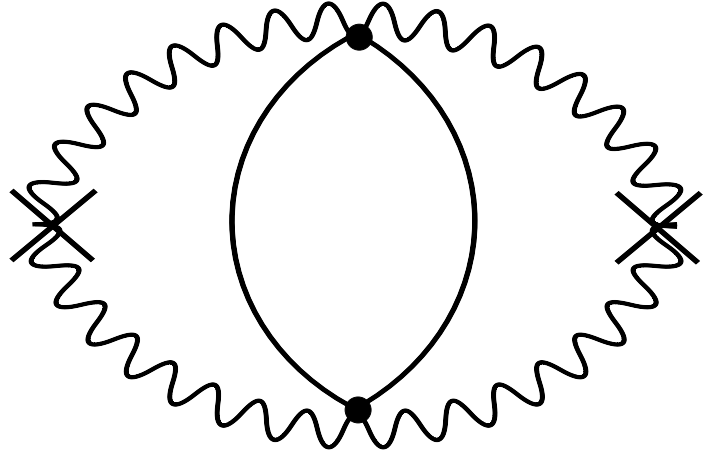}
\end{center}
\caption{Vanishing irreducible three-loop diagram for the correlator
  $\bra\Theta_{00}(x)\Theta_{00}(y)\ket$ in a thermalized, deconfining SU(2) gauge
theory. A wavy (solid) line is associated with
the propagator of the massless (massive) mode.
Crosses denote the insertion of the composite operator $\Theta_{00}$.\label{Fig-3a}}
\end{figure}

\begin{figure}
\begin{center}
\vspace{5.3cm}
\includegraphics{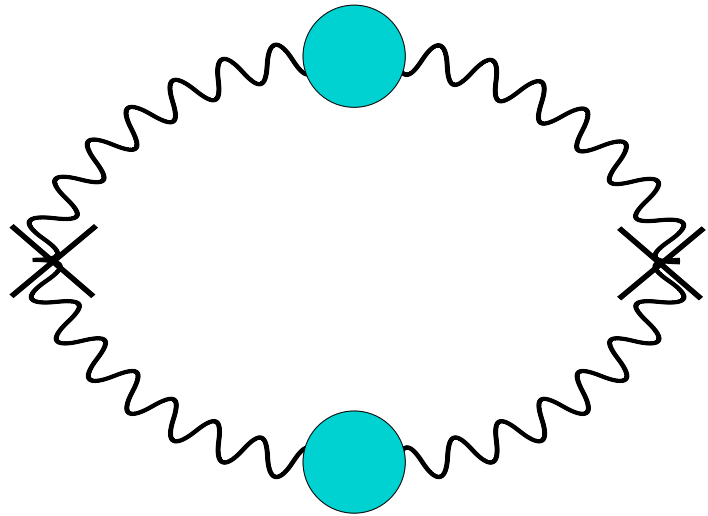}
\end{center}
\caption{Feynman diagram for the correlator
  $\bra\Theta_{00}(x)\Theta_{00}(y)\ket$ in a thermalized, deconfining SU(2) gauge
theory including radiative corrections to lowest order. (Blobs signal a
resummation of the one-loop polarization for the massless mode.)
Crosses denote the insertion of the composite operator $\Theta_{00}$.\label{Fig-4}}
\end{figure}

In \cite{SHG2006-1} the one-loop polarization tensor $\Pi_{\mu\nu}$ for
the on-shell ($p^2=0$)
massless mode $A^3_\mu$ was computed. As a result, a modification of the
dispersion law
\eqb
\label{displawmod}
p_0^2=\vec{p}^2 \to p_0^2=\vec{p}^2+G(T,|\vec{p}|,\Lambda)
\eqe
was obtained where $\Lambda$
denotes the Yang-Mills scale related to the critical temperature $T_c$
for the deconfining-preconfining phase transition as
$T_c=\frac{\lambda_c}{2\pi}\,\Lambda=\frac{13.87}{2\pi}\,\Lambda$. For temperatures not much larger than
$T_c$ the function $G$ acquires relevance: There is a regime of
antiscreening ($G<0$) for spatial momenta larger than
$|\vec{p}_{\tiny\mbox{as}}|\sim 0.2\,T$. This effect, however, dies off
exponentially fast with increasing momenta. For momenta smaller than
$|\vec{p}_{\tiny\mbox{high}}|\sim 0.1\,T$ and larger than
$|\vec{p}_{\tiny\mbox{low}}|\sim 0.02\,T$ the function $G$ is so strongly
positive that the propagation of the associated modes
is forbidden (total screening). The situation is summarized in
Fig.\,\ref{Fig-3} where the logarithm of $\frac{G}{T^2}$ is
plotted as a function of dimensionless spatial momentum modulus
$|\tilde{\vec{p}}|$ and for various temperature not to far above
$T_c$. (The dimensionless temperature $\lambda$ is defined as $\lambda\equiv
13.87\frac{T}{T_c}=\frac{2\pi T}{\Lambda}$.) The function
$f(|\tilde{\vec{p}}|)\equiv 2\log_{10}|\tilde{\vec{p}}|$
marks the line at which the screening mass of a photon equals the
modulus of its spatial momentum.
It is sufficient to account for radiative
corrections in the correlator
$\bra\Theta_{00}(\vec{x})\Theta_{00}(\vec{y})\ket^{\tiny\mbox{th}}$ in
terms of a resummation of the one-loop polarization tensor for the
massless mode only, see Fig.\,\ref{Fig-4}, since the one irreducible three-loop
diagram vanishes identically \cite{KH2007}, see Fig.\,\ref{Fig-3a}, and the other,
containing a two-loop irreducible contribution to the polarization, is
entirely negligible \cite{KH2007}.

\subsection{Thermal part of two-point correlation}

In analogy to Eq.\,(\ref{thermcalcu1}) the thermal part
$\bra\Theta_{00}(\textbf{x})\Theta_{00}(\textbf{y})\ket^{\tiny\mbox{th}}$
now calculates as
\begin{eqnarray}
\label{rohthetather}
\bra\Theta_{00}(\vec{x})\Theta_{00}(\vec{y})\ket^{\tiny\mbox{th}}&=&
\frac{1}{2}\left(\int\!\!\frac{d^3p}{(2\pi)^3}
        \sqrt{\vec{p}^2+G}
        \,n_B(\beta\sqrt{\vec{p}^2+G})
        \,\e^{i\vec{p}(\vec{x}-\vec{y})}\right)^2\nonumber\\
& &+\frac{1}{2}\left(\int\!\!\frac{d^3p}{(2\pi)^3}
        \frac{\vec{p}^2}{\sqrt{\vec{p}^2+G}}
        \,n_B(\beta\sqrt{\vec{p}^2+G})
        \,\e^{i\vec{p}(\vec{x}-\vec{y})}\right)^2\nonumber\\
& &+\frac{1}{2}\int\!\!\frac{d^3p}{(2\pi)^3}\int\!\!\frac{d^3k}{(2\pi)^3}
        \left(\frac{\vec{p}\vec{k}}{|\vec{p}||\vec{k}|}\right)^2
        \sqrt{\vec{p}^2+G}\sqrt{\vec{k}^2+G}\times\nonumber\\
       & & \ \ \ \ \ \ n_B(\beta\sqrt{\vec{p}^2+G})n_B(\beta\sqrt{\vec{k}^2+G})
        \,\e^{i\vec{p}(\vec{x}-\vec{y})}\e^{i\vec{k}(\vec{x}-\vec{y})}\nonumber\\
& &+\frac{1}{2}\int\!\!\frac{d^3p}{(2\pi)^3}\int\!\!\frac{d^3k}{(2\pi)^3}
        \left(\frac{\vec{p}\vec{k}}{|\vec{p}||\vec{k}|}\right)^2
        \frac{\vec{p}^2}{\sqrt{\vec{p}^2+G}}\frac{\vec{k}^2}{\sqrt{\vec{k}^2+G}}\times\nonumber\\
        & &\ \ \ \ \ \ n_B(\beta\sqrt{\vec{p}^2+G})n_B(\beta\sqrt{\vec{k}^2+G})
        \,\e^{i\vec{p}(\vec{x}-\vec{y})}\e^{i\vec{k}(\vec{x}-\vec{y})}\,.\nonumber\\
\end{eqnarray}

The integration over the $p_0$- and $k_0$-coordinates is performed
without constraints and
just fixes the dispersion law (\ref{displawmod}). As in the U(1) case we
introduce spherical coordinates and evaluate
the integrals over the angles analytically. Upon performing the azimuthal integrations for
each summand in Eq.\,(\ref{rohthetather}) the respective expression
reduces to a square of an integral over the modulus of spatial momentum. This integral is
treated in analogy
to the derivation of the black-body spectrum in \cite{SHG2006-2}.
Employing the modified dispersion law, the integral over
momentum-modulus is replaced in favor of an integral over frequency
$\omega$. Those values of $\omega$, which yield an
imaginary modulus of the spatial momentum due to strong screening, are excluded from the domain
of integration. This regime of strong screening is in the range
$\omega_1<\omega<\omega_2$, where $\omega_1, \omega_2$ denote the
solutions of the equation
\eqb
\label{roots}
\omega^2-G(\omega,T,\Lambda)=0\,.
\eqe
\newpage
Finally, we
introduce a dimensionless frequency $\tilde{\omega}\equiv\beta\omega$
and a dimensionless screening function $\tilde{G}\equiv\beta^2 G$ and arrive at
\begin{eqnarray}
\label{thermsu2}
&&\bra\Theta_{00}(\vec{x})\Theta_{00}(\vec{y})\ket^{\tiny\mbox{th}}=\nonumber\\
&&\frac{1}{(2\pi)^6\beta^8}\left(\frac{32\pi^2}{3}
    \left(\int\limits_{0\leq\tilde{\omega}\leq\tilde{\omega}_1,
    \atop\tilde{\omega}_2\leq\tilde{\omega}\leq\infty}
    d\tilde{\omega}\left(\tilde{\omega} -\frac{1}{2}\frac{d\tilde{G}}{d\tilde{\omega}}\right)\,\tilde{\omega}
    \sqrt{\tilde{\omega}^2-\tilde{G}}\,
    \frac{j_0(\sqrt{\tilde{\omega}^2-\tilde{G}}\,|\tilde{\vec{z}}|)}
    {e^{\tilde{\omega}}-1}\right)^2\right.\nonumber\\
&&\left.+\,\frac{32\pi^2}{3}
    \left(\int\limits_{0\leq\tilde{\omega}\leq\tilde{\omega}_1,
    \atop\tilde{\omega}_2\leq\tilde{\omega}\leq\infty}
    d\tilde{\omega}\left(\tilde{\omega}
    -\frac{1}{2}\frac{d\tilde{G}}{d\tilde{\omega}}\right)\,
    \frac{(\sqrt{\tilde{\omega}^2-\tilde{G}})^3}{\tilde{\omega}}\,
    \frac{j_0(\sqrt{\tilde{\omega}^2-\tilde{G}}\,|\tilde{\vec{z}}|)}
    {e^{\tilde{\omega}}-1}\right)^2\right.\nonumber\\
&&\left.+\,\frac{16\pi^2}{3}\left(\int\limits_{0\leq\tilde{\omega}\leq\tilde{\omega}_1,
    \atop\tilde{\omega}_2\leq\tilde{\omega}\leq\infty}
    d\tilde{\omega}\left(\tilde{\omega}
    -\frac{1}{2}\frac{d\tilde{G}}{d\tilde{\omega}}\right)\,\,
    \tilde{\omega}\sqrt{\tilde{\omega}^2-\tilde{G}}\,
    \frac{j_2(\sqrt{\tilde{\omega}^2-\tilde{G}}\,|\tilde{\vec{z}}|)}
    {e^{\tilde{\omega}}-1}\right)^2\right.\nonumber\\
&&\left.+\,\frac{16\pi^2}{3}\left(\int\limits_{0\leq\tilde{\omega}\leq\tilde{\omega}_1,
    \atop\tilde{\omega}_2\leq\tilde{\omega}\leq\infty}
    d\tilde{\omega}\left(\tilde{\omega} -\frac{1}{2}\frac{d\tilde{G}}{d\tilde{\omega}}\right)\,
    \frac{(\sqrt{\tilde{\omega}^2-\tilde{G}})^3}{\tilde{\omega}}\,
    \frac{j_2(\sqrt{\tilde{\omega}^2-\tilde{G}}\,|\tilde{\vec{z}}|)}
    {e^{\tilde{\omega}}-1}\right)^2\right)\,.\nonumber\\
\end{eqnarray}

\subsection{Estimate for vacuum  part of two-point correlation}

Here we would like to obtain an order-of-magnitude estimate
for the vacuum part of the two-point correlation of $\Theta_{00}$
(massless mode) in deconfining SU(2) Yang-Mills thermodynamics. To do this we
ignore the modification of the dispersion law in
Eq.\,(\ref{displawmod}). Anyways, the function $G$ has so far only been
computed for external momentum $p$ with $p^2=0$. The difference as
compared to the U(1) case is then a restriction of the (euclidean)
four-momentum $p$ as $p^2\le|\phi|^2$ due to the existence
of a scale of maximal resolution $|\phi|$ in the effective
theory. Again, we will see that
$\bra\Theta_{00}(\vec{x})\Theta_{00}(\vec{y})\ket^{\tiny\mbox{vac}}$ is
a negligible correction to
$\bra\Theta_{00}(\vec{x})\Theta_{00}(\vec{y})\ket^{\tiny\mbox{th}}$
for physically interesting distances.

In case of
$\bra\Theta_{00}(x)\Theta_{00}(y)\ket^{\tiny\mbox{vac}}$ we
again restrict to $x_0=y_0$ to be able to compare with
$\bra\Theta_{00}(\textbf{x})\Theta_{00}(\textbf{y})\ket^{\tiny\mbox{th}}$.
Introducing the dimensionless version of $\phi$ as $\tilde{\phi}\equiv\beta\phi$, proceeding in a
way analogous to the derivation of Eq.\,(\ref{before2polvacu1}), and
performing the $\psi$- and $|\tilde{p}|$-integrations, we arrive at
\begin{eqnarray}
\label{vacsu2}
&&\bra\Theta_{00}(\vec{x})\Theta_{00}(\vec{y})\ket^{\tiny\mbox{vac}}
   \sim\nonumber\\
&&\frac{1}{(2\pi)^8\beta^8}\Bigg(96\pi^2
   \cdot\frac{\pi^2}{|\vec{\tilde{z}}|^8}
   \left(2J_0(|\vec{\tilde{z}}||\tilde{\phi}|)
   +|\vec{\tilde{z}}||\tilde{\phi}|J_1(|\vec{\tilde{z}}||\tilde{\phi}|)
   -2\right)^2 \nonumber\\
&+&\left.48\pi^2\cdot\frac{\pi^2}{4|\vec{\tilde{z}}|^8}
   \left(2(8J_0(|\vec{\tilde{z}}||\tilde{\phi}|)+
   |\vec{\tilde{z}}||\tilde{\phi}|J_1(|\vec{\tilde{z}}||\tilde{\phi}|)-8)
   +3|\vec{\tilde{z}}|^2|\tilde{\phi}|^2\,\,_1F_2\left(\frac{1}{2}
   ;\frac{3}{2},2;-\frac{|\vec{\tilde{z}}|^2|\tilde{\phi}|^2}{4}\right)
   \right)^2\right.\nonumber\\
&+&\left.\frac{32\pi^2}{3}\cdot\frac{\pi^2}{|\vec{\tilde{z}}|^8}
   \left((|\vec{\tilde{z}}|^2|\tilde{\phi}|^2-2)
   J_0(|\vec{\tilde{z}}||\tilde{\phi}|)-3|\vec{\tilde{z}}||\tilde{\phi}|
   J_1(|\vec{\tilde{z}}||\tilde{\phi}|)+2\right)^2\right.  \nonumber\\
&+&\left.\frac{16\pi^2}{3}\cdot\frac{\pi^2}{|\vec{\tilde{z}}|^8}
   \left((|\vec{\tilde{z}}|^2|\tilde{\phi}|^2-8)
   J_0(|\vec{\tilde{z}}||\tilde{\phi}|)
   -6|\vec{\tilde{z}}||\tilde{\phi}|J_1(|\vec{\tilde{z}}||\tilde{\phi}|)
   +8\right)^2\right.  \nonumber\\
&+&\left.32\pi^2\cdot\frac{\pi^2|\tilde{\phi}|^8}{64}
   \left(\,\,_1F_2\left(\frac{1}{2};\frac{3}{2},3;
   -\frac{|\vec{\tilde{z}}|^2|\tilde{\phi}|^2}{4}\right)\right)^2\right.\nonumber\\
&+&\left.24\pi^2\cdot\frac{\pi^2|\tilde{\phi}|^4}
   {|\vec{\tilde{z}}|^4}\left(J_2(|\vec{\tilde{z}}||\tilde{\phi}|)\right)^2
   \right.  \nonumber\\
&+&48\pi^2\cdot\frac{\pi^2|\tilde{\phi}|^2}
   {4|\vec{\tilde{z}}|^6}\left(6J_1(|\vec{\tilde{z}}||\tilde{\phi}|)+
   2|\vec{\tilde{z}}||\tilde{\phi}|
   J_2(|\vec{\tilde{z}}||\tilde{\phi}|)-3|\vec{\tilde{z}}||\tilde{\phi}|
   \,\,_1F_2\left(\frac{1}{2};\frac{3}{2},2;
   -\frac{|\vec{\tilde{z}}|^2|\tilde{\phi}|^2}{4}\right)\right)^2\Bigg)
   \,,\nonumber\\
\end{eqnarray}
where $\,_1F_2$ is a hypergeometric function and $J_0, J_1, J_2$ are Bessel
functions of the first kind (conventions as in \cite{Gradshteyn}).

\subsection{Numerical results\label{numsec}}

\begin{figure}
\begin{center}
\vspace{5.3cm}
\includegraphics{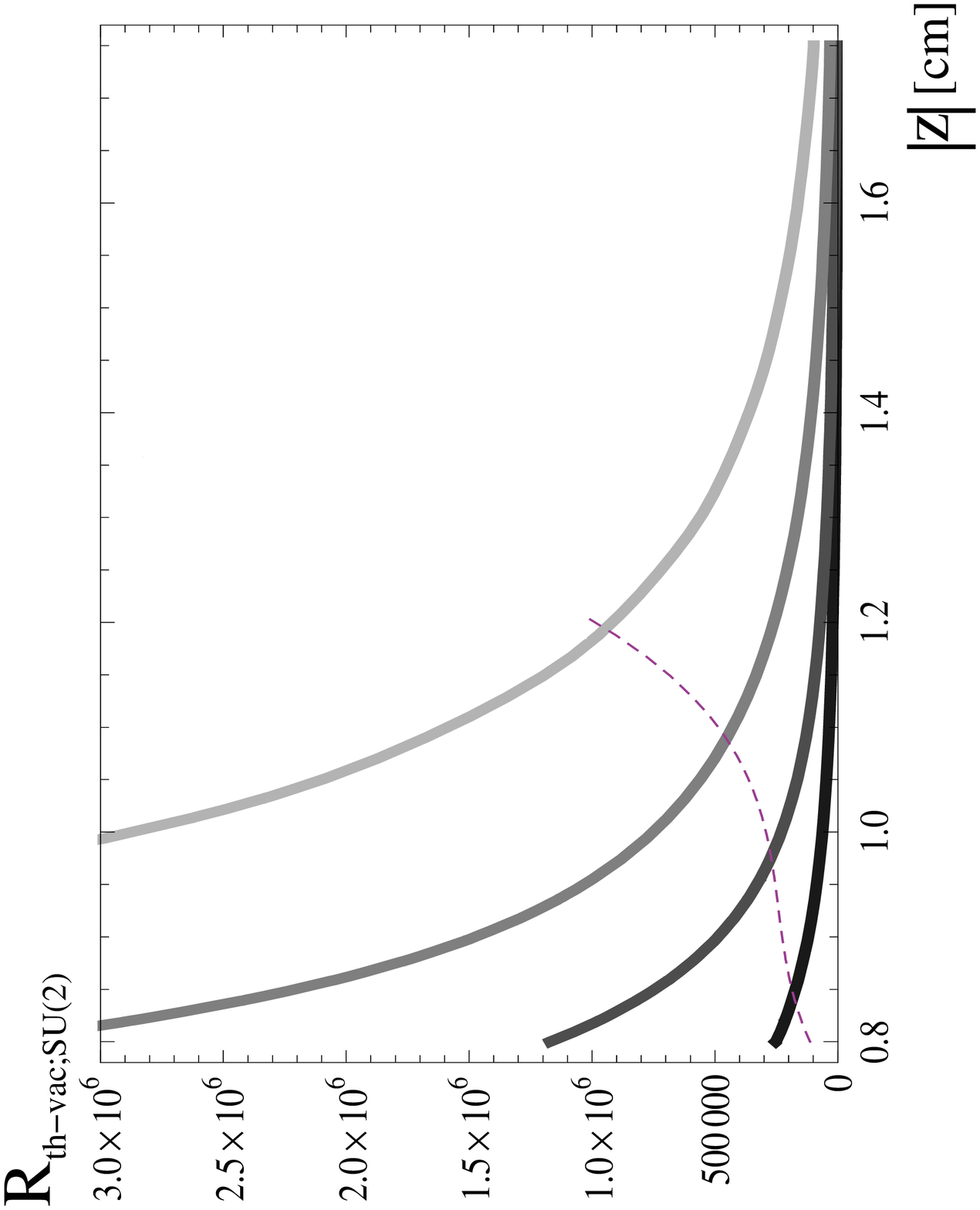}
\end{center}
\caption{\protect{\label{Fig-5}The function
  $R_{\tiny\mbox{th}-\tiny\mbox{vac};\tiny\mbox{SU(2)}}(|\vec{z}|)$,
  defined as in Eq.\,(\ref{defRsu2}), when specializing to
the case of SU(2)$_{\tiny\mbox{CMB}}$ ($T_c=2.73\,$K), for
various temperatures: black curve ($T=1.5\,T_c$), dark grey curve
($T=2.0\,T_c$), grey curve ($T=2.5\,T_c$), light grey curve ($T=3.0\,T_c$). The dashed line separates distances
smaller than $|\phi|^{-1}$ from those that are larger than $|\phi|^{-1}$.}}
\end{figure}

The integral in Eqs.\,(\ref{thermsu2}) is evaluated numerically. Let us
first compare the estimate for the vacuum contribution in Eq.\,(\ref{vacsu2}) with the thermal
part of the correlator
$\bra\Theta_{00}(\vec{x})\Theta_{00}(\vec{y})\ket$. To make contact
with SU(2)$_{\tiny\mbox{CMB}}$, whose Yang-Mills scale is
$\Lambda=1.065\times10^{-4}$\,eV \cite{SHG2006-1,SHG2006-2},
we relate at a given temperature the
dimensionless distance $|\tilde{\vec{z}}|$ to the physical distance in
centimeters. We define
\eqb
\label{defRsu2}
R_{\tiny\mbox{th}-\tiny\mbox{vac};\tiny\mbox{SU(2)}}(|\vec{z}|)\equiv
\frac{\bra\Theta_{00}(\vec{x})\Theta_{00}(\vec{y})\ket^{\tiny\mbox{th}}}
{\left|\bra\Theta_{00}(\vec{x})\Theta_{00}(\vec{y})\ket^{\tiny\mbox{vac}}\right|}\,,
\eqe
where we use the estimate in Eq.\,(\ref{vacsu2}) for
$\left|\bra\Theta_{00}(\vec{x})\Theta_{00}(\vec{y})\ket^{\tiny\mbox{vac}}\right|$.
In Fig.\,\ref{Fig-5} the quantity
$R_{\tiny\mbox{th}-\tiny\mbox{vac};\tiny\mbox{SU(2)}}(|\vec{z}|)$ is
depicted for various temperatures specializing to the case
of SU(2)$_{\tiny\mbox{CMB}}$. Here we have used the one-loop selfconsistent determination of the function $G$ such that 
$G=p^2$, see \cite{LH2008}. The thus obtained improvement of $G$ shows that $\tilde{\omega}_1\equiv 0$ in Eq.\,(\ref{thermsu2}).

Notice the strong dominance of the
thermal part. Notice also that although this result resembles
qualitatively the result of Fig.\,\ref{Fig-2} for fixed distance and
varying temperature this is not true for fixed temperature and varying
distance. Namely, the existence of a nontrivial thermal ground
state in deconfining SU(2) Yang-Mills thermodynamics constrains quantum
fluctuations of the massless mode to be softer than the scale
$|\phi|$. Owing to its trivial ground state,
no such constraint exists in a thermalized U(1) gauge theory. As a
consequence and in accord with Fig.\,\ref{Fig-2}, quantum fluctuations dominate thermal
fluctuations at small distances in such a theory.

\begin{figure}
\begin{center}
\vspace{5.3cm}
\includegraphics{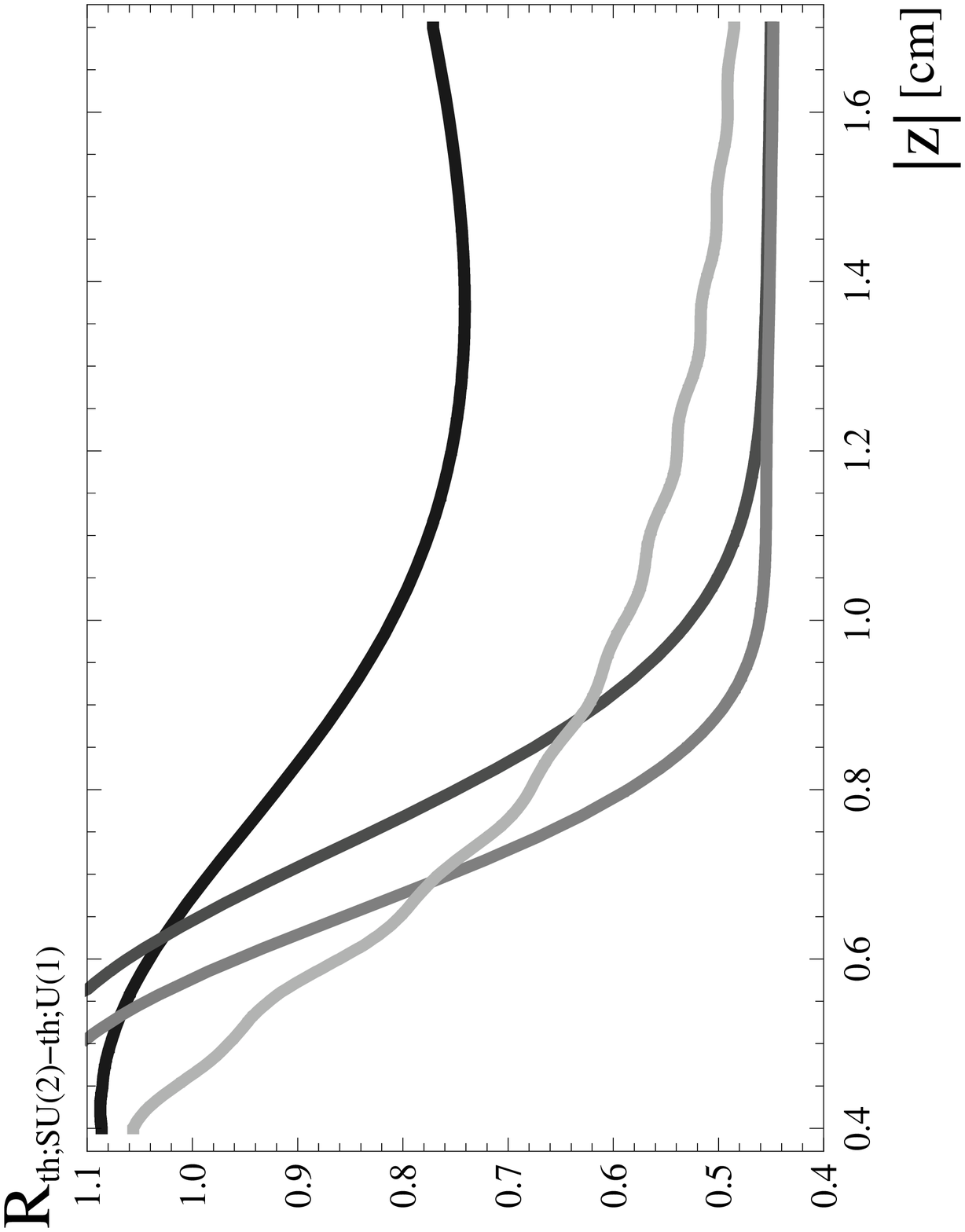}
\end{center}
\caption{\protect{\label{Fig-6}The function
  $R_{\tiny\mbox{th,SU(2)}-\tiny\mbox{th,U(1)}}(|\vec{z}|)$,
as defined in Eq.\,(\ref{defRsu2u1}), when specializing to
the case of SU(2)$_{\tiny\mbox{CMB}}$ ($T_c=2.73\,$K), for
various temperatures: black curve ($T=1.5\,T_c$), dark grey
($T=2.0\,T_c$), grey ($T=2.5\,T_c$), light grey ($T=3.0\,T_c$). Notice the regime of antiscreening for small 
$|z|$}. For $T=3.0\,T_c$, where the approximation $p^2=0$ deviates sizably from the full result for the screening function $G$, 
we have used $G$ as obtained 1-loop selfconsistently \cite{LH2008}.}
\end{figure}

By virtue of the results shown in
Fig.\,\ref{Fig-2} and Fig.\,\ref{Fig-5}
we neglect the vacuum contribution to
$\bra\Theta_{00}(\vec{x})\Theta_{00}(\vec{y})\ket$ in the following.
Let us now turn to the interesting question of how much suppression
there is in the correlation of the photon energy density
in the case of SU(2)$_{\tiny\mbox{CMB}}$ as compared to the conventional
U(1) case. In Fig.\,\ref{Fig-6} the ratio
$R_{\tiny\mbox{th,SU(2)}-\tiny\mbox{th,U(1)}}(|\vec{z}|)$,
defined as
\eqb
\label{defRsu2u1}
R_{\tiny\mbox{th,SU(2)}-\tiny\mbox{th,U(1)}}(|\vec{z}|)\equiv
\frac{\bra\Theta_{00}(\vec{x})\Theta_{00}(\vec{y})\ket^{\tiny\mbox{th,SU(2)}}}
{\bra\Theta_{00}(\vec{x})\Theta_{00}(\vec{y})\ket^{\tiny\mbox{th,U(1)}}}\,,
\eqe
is depicted for various temperatures as a function of distance in
centimeters. Again we have used the one-loop selfconsistent determination of the function $G$ such that 
$G=p^2$, see \cite{LH2008}.  
Notice the suppression of the correlation between photon energy
densities in the case of an underlying SU(2) gauge symmetry as compared
to the case of the conventional U(1). Since the correlator
$\bra\Theta_{00}(\vec{x})\Theta_{00}(\vec{y})\ket$
is a measure for the energy transfer between the spatial points
$\vec{x}$ and $\vec{y}$ in thermal equilibrium and hence a measure
for the interaction of microscopic objects emitting and absorbing
radiation, we conclude that this interaction is, as compared to the
conventional theory, suppressed on distances
$\sim 1\,$cm if photon propagation is subject to an SU(2) gauge
principle.

\section{Stability of clouds of
atomic hydrogen in the Milky Way\label{MW}}

The results of Sec.\,\ref{decsu2} have immediate implications for the
gradual metamorphosis of cold
($T=5\dots 10\,$K) astrophysical objects such as the
hydrogen cloud GSH139-03-69 observed in between spiral arms of the outer
Milky Way \cite{BruntKnee2001}. This object possesses an estimated age of about
50 million years, exhibits a brightness temperature $T_B$ of $T_B\sim 20\,$K with
cold regions of $T_B\sim 5\cdots 10\,$K and an atomic number density of
$\sim 1.5\,$cm$^{-3}$. The puzzle about this and similar structures
seems to be its high content of atomic hydrogen in view of
its unexpectedly large age. Namely, numerical simulations of the
cloud evolution subject to standard interatomic forces suggest a much
lower time scale of less than 10 million years, see \cite{GoldsmithI,GoldsmithII,GoldsmithIII} and
references therein, for the generation of a
substantial fraction of H$_2$ molecules.

Although a quantitative estimate of the
increased stability of atomic hydrogen clouds due
to the SU(2) effects in thermalized photon propagation
is beyond the scope of the present work,
Fig.\,\ref{Fig-6} clearly expresses that the mean energy
transfer in the photon gas of $T=5\cdots 10\,$K is suppressed by up to a
factor of two at the interatomic distances in
the hydrogen cloud GSH139-03-69. This, however, implies
that atomic interactions, leading to the formation of molecules,
are strongly suppressed. It would be interesting to see how the
simulation of the cloud evolution, taking into account the effects as
expressed in Fig.\,\ref{Fig-6}, would increase the estimate for its age
as compared to the standard picture.

As already pointed out in \cite{SHG2006-2}, the propagation of the 21-cm
line is unscreened even when
subjecting photons to SU(2)$_{\tiny\mbox{CMB}}$. However, a recent full
calculation of $G$ shows that this result is an artefact of
the approximation $p^2=0$ \cite{LH2008}: Thermalization of the hydrogen cloud
thus takes place solely via the coupling of the photon to the nontrivial
thermal ground state \cite{GHKL2008}.

\section{Summary and Conclusions\label{C}}

In this work we have computed the two-point correlation of the canonical
energy density of photons both in the conventional and
postulated case of a U(1) and SU(2) gauge group, respectively.
In the real-time formalism of finite-temperature field theory
and resumming the polarization tensor for the
massless mode (photon) \cite{SHG2006-1} in the SU(2) case, this correlation splits into a thermal and a
vacuum part\footnote{Higher than one-loop
irreducible diagrams vanish identically \cite{KH2007}.}.
We have observed that for the case of SU(2)$_{\tiny\mbox{CMB}}$,
which is postulated to be the nonabelian gauge theory
underlying photon propagation \cite{HofmannB2005,SHG2006-2,GH2005},
the vacuum part is strongly suppressed as compared
to the thermal part for distances $\sim 1$\,cm.
Furthermore, there is strong suppression of the
SU(2)$_{\tiny\mbox{CMB}}$-correlator as compared
to the U(1)-correlator at the interatomic distances of hydrogen
atoms within unexpectedly stable and cold ($T\sim 5\cdots 10\,$K)
cloud structures in between spiral arms of the Milky Way. Thus the mean
energy transport and hence the atomic interactions are hamstrung in such
structures possibly explaining their unsuspectedly large age.

\begin{figure}
\begin{center}
\vspace{5.3cm}
\includegraphics{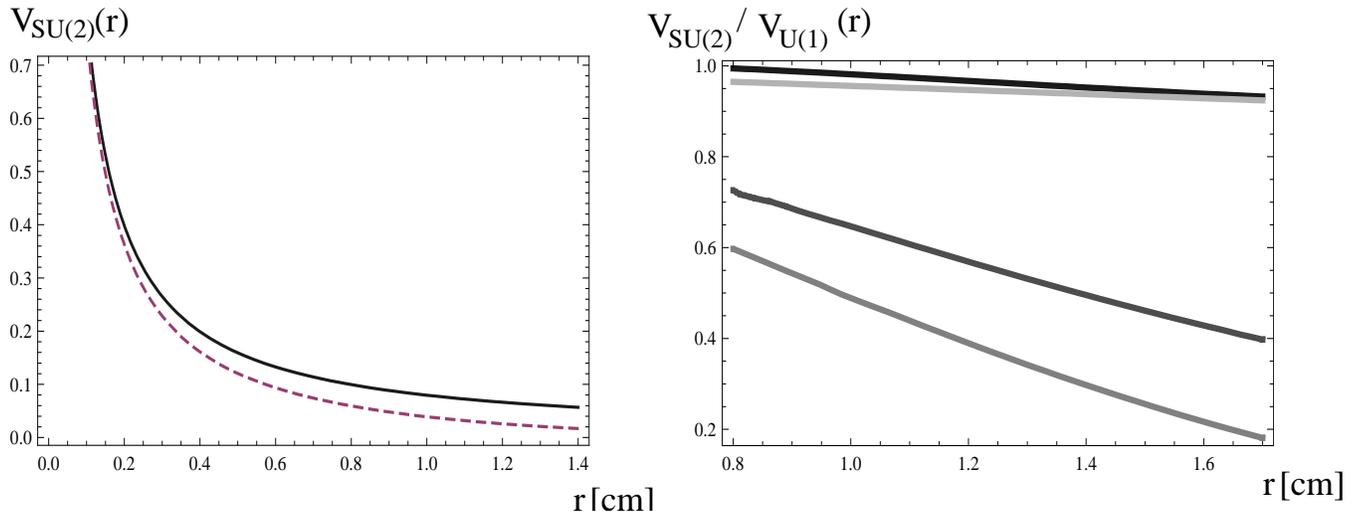}
\end{center}
\caption{\protect{\label{Fig-7} Left panel: Plot of the potentials
$V_{\tiny\mbox{U(1)}}(r)$ (dashed line) and
$V_{\tiny\mbox{SU(2)}_{\tiny\mbox{CMB}}}(r)$ (solid line)
as a function of $r$ at $T=2.5\,T_{\tiny\mbox{CMB}}\sim 6.8$\,K. Right panel: Plot of the
ratio
$V_{\tiny\mbox{SU(2)}_{\tiny\mbox{CMB}}}(r)/V_{\tiny\mbox{U(1)}}(r)$ as
a function of $r$. The temperature is set to $T=1.5\,T_c$ (black),
$T=2.0\,T_c$ (dark grey), $T=2.5\,T_c$ (grey), and $T=3.0\,T_c$ (light
grey). For $T=3.0\,T_c$, where the approximation $p^2=0$ deviates sizably from the full result for the screening function $G$, 
we have used $G$ as obtained 1-loop selfconsistently \cite{LH2008}. Notice how close to unity the curve at $T=3.0\,T_c$ 
is as compared to the curve at $T=2.5\,T_c$ for the right-hand side panel.}}
\end{figure}

To make
the situation even more explicit we have computed the
Coulomb potential $V(r)$ ($r\equiv|\vec{x}|$) of a heavy point charge in case of photons being
described by a U(1) and an SU(2) gauge theory. This potential is given in a U(1)
theory as
\eab
\label{CU1}
V_{\tiny\mbox{U(1)}}(r)&=&\frac{1}{(2\pi)^3}\int
d^3p\,\frac{\e^{-i\vec{p}\cdot\vec{x}}}{\vec{p}^2}=\frac{1}{2\pi^2}\int_0^\infty
dp\,\frac{\sin pr}{pr}=\frac{1}{2\pi^2r}\int_0^\infty
d\xi\,\frac{\sin\xi}{\xi}\nonumber\\
&=&\frac{1}{4\pi r}\,.
\eae
Going from U(1) to SU(2), we take into
account the resummed one-loop polarization \cite{SHG2006-1}
by letting $\vec{p}^2\to\vec{p}^2+G$ in the denominator of the integrand
in Eq.\,(\ref{CU1}). Here $G$ is the same function as discussed in
Sec.\,\ref{RMd}. The physical situation is a heavy point charge
immersed into the SU(2) plasma with the photons associated with it being
(anti)screened by nonabelian, thermal fluctuations. Although the
function $G$ is known for on-shell photons $\omega^2=\vec{p}^2$
only this recipe should work well for sufficiently large distances since the
off-shellness of the photon, which does not mediate any energy transfer
from the source, then is sufficiently small. Thus for
SU(2) we approximately\footnote{A point charge immersed into the plasma
  locally distorts the latter's ground state, and, strictly speaking, the
theory to describe this distortion yet needs to be worked out.
But measuring the (anti)screening of the potential sufficiently far away from the location of
the charge should still be describable in terms
of unadulturated SU(2) Yang-Mills thermodynamics.}
have
\eab
\label{SU2}
V_{\tiny\mbox{SU(2)}}(r)&=&\frac{1}{(2\pi^3)}\int
d^3p\,\frac{\e^{-i\vec{p}\cdot\vec{x}}}{\vec{p}^2+G(T,|\vec{p}|,\Lambda)}\nonumber\\
&=&\frac{1}{2\pi^2r}\int_0^\infty
dp\,\frac{p}{p^2+G(T,p,\Lambda)}\sin pr\,,
\eae
where the last integral is performed numerically. In Fig.\,\ref{Fig-7} both
potentials, $V_{\tiny\mbox{U(1)}}(r)$ and $V_{\tiny\mbox{SU(2)}_{\tiny\mbox{CMB}}}(r)$, as
well as their ratio are plotted as functions of $r$.
Notice that the dimensionless quantity
$V_{\tiny\mbox{SU(2)}_{\tiny\mbox{CMB}}}(r)/V_{\tiny\mbox{U(1)}}(r)$
shows similar suppression as the dimensionless quantity
$R_{\tiny\mbox{th,SU(2)}-\tiny\mbox{th,U(1)}}(|\vec{z}|)$ depicted in
Fig.\,\ref{Fig-6}. This seems to confirm the validity of
our above approximation. However, we wish to stress that such a simple
approximation cannot be used in a general framework: Apart from the partial
knowledge of the function $G$ (known on-shell only) our thermodynamical
approach to SU(2)$_{\tiny\mbox{CMB}}$ and its application to the
screening of Coulomb potnetials breaks down when macroscopic sources
with large multiples of the elementary electric charge are considered. In fact, in this
case large energy densities ($\rho\gg T^4$) are locally present which destroy the
applicability of a thermalized theory. But the effective, Debye massive nature
of the photon can only be detected in weakly adulterated thermal
systems (black body) \cite{SHG2006-2} and is relevant in the cold atomic
hydrogen clouds as discussed in this work.

\section*{Acknowledgments}

The authors would like to thank Markus Schwarz for useful conversations and his
helpful comments on the manuscript. We are also grateful for productive
criticism by a Referee.

\end{document}